\documentclass[%
showpacs,
bibnotes,
amsmath,amssymb,
pra,						     
twocolumn,                   
letterpaper,
twoside,
frontmatterverbose,
floatfix,
notitlepage,
]{revtex4-1}

\usepackage{graphicx}
\usepackage{dcolumn}
\usepackage{bm}
\usepackage{hyperref}
\usepackage{multirow}
\usepackage{array}
\usepackage{booktabs}
\usepackage{ctable}
\usepackage{upgreek}
\usepackage{epsfig}
\usepackage{mathrsfs}
\usepackage{amssymb}
\usepackage{amsbsy}
\usepackage{color}
\usepackage{cancel}
\usepackage{pifont}
\usepackage{marginnote}
\usepackage{float}
\usepackage{tikz}
\usetikzlibrary{arrows}
\tikzstyle{block}=[draw opacity=0.7,line width=1.4cm]
\usetikzlibrary{positioning}
\usepackage[caption=false]{subfig}
\usepackage{setspace}

\newcommand{\bcen}{\begin{center}}
\newcommand{\ecen}{\end{center}}
\newcommand{\btab}{\begin{tabular}}
\newcommand{\etab}{\end{tabular}}
\newcommand{\bdes}{\begin{description}}
\newcommand{\edes}{\end{description}}

\newcommand{\beq}{\begin{equation}}
\newcommand{\eeq}{\end{equation}}
\newcommand{\bea}{\begin{eqnarray}}
\newcommand{\eea}{\end{eqnarray}}

\newcommand{\half}{\frac{1}{2}}
\newcommand{\bary}{\begin{array}}
\newcommand{\eary}{\end{array}}
\newcommand{\benum}{\begin{enumerate}}
\newcommand{\eenum}{\end{enumerate}}
\newcommand{\bitem}{\begin{itemize}}
\newcommand{\eitem}{\end{itemize}}

\newcommand{\blam}{{\boldsymbol{\lambda}}}

\newcommand{\be} { \mbox{\boldmath $e$}}

\newcommand{\bk} { \bm{k} }

\newcommand{\mean}[1]{\langle #1 \rangle}

\newcommand{\paratitle}[1]{\noindent {\bf #1:}}

\newcommand{\eqn}[1] {eqn.~(\ref{#1})}

\newcommand{\sect}[1] {Section~\ref{#1}}

\newcommand{\fig}[1]{fig.~\ref{#1}}
\newcommand{\Fig}[1]{Fig.~\ref{#1}}

\makeatletter

\newcommand{\Rmnum}[1]{\expandafter\@slowromancap\romannumeral #1@}
\makeatother

\newcommand{\myfigwidth}{0.99\columnwidth}

\newcommand{\ie}{{i.e., } }

\newcommand{\mylabel}[1]{\label{#1}} 

\newcommand{\mycite}[1]{\cite{#1}}

\newcommand{\authorprl}[2]{\author{#1}\email{#2}}

\newcommand{\titlename}{Fractional Local Moment and High Temperature Kondo Effect in Rashba-Fermi Gases}

\begin{document}


\title{\titlename}
\authorprl{Adhip Agarwala}{adhip@physics.iisc.ernet.in}
\authorprl{Vijay B. Shenoy}{shenoy@physics.iisc.ernet.in}
\affiliation{Centre for Condensed Matter Theory, Department of Physics, Indian Institute of Science, Bangalore 560 012, India}


\date{\today}


\begin{abstract} 
 We investigate the new physics that arises when a correlated quantum impurity hybridizes with Fermi gas with a generalized Rashba spin-orbit coupling produced via a uniform synthetic non-Abelian gauge field. We show that the impurity develops a {\em fractional} local moment which couples anti-ferromagnetically to the Rashba-Fermi gas. This results in a concomitant  {\em Kondo effect with a high temperature scale} that can be tuned by the strength of the Rashba spin-orbit coupling.
\end{abstract}

\maketitle

Quantum emulation of many particle systems\mycite{Ketterle_arXiv_2008, Bloch_RMP_2008, Esslinger_ARCMP_2010, Bloch_NatPhys_2012, Cirac_NatPhys_2012} with cold atoms can not only help address key open problems across physics disciplines, but also explore phenomena in regimes and conditions not accessible in conventional systems. Experimental progress in cold atoms over the last decade has provided insights into several ``classic'' systems such as the Bose-Hubbard model\mycite{Jaksch_PRL_1998,Greiner_Nature_2002} and the BCS-BEC crossover of fermions\mycite{Regal_PRL_2004, Zwierlein_PRL_2004}. The scope of cold atoms has been significantly enhanced by the recent realizations of systems with synthetic gauge fields\mycite{Lin_Nature_2011, Williams_Science_2012, Wang_PRL_2012, Cheuk_PRL_2012,Huang_arXiv_2015} (see \mycite{Goldman_RPP_2014} for a review), and has even lead to realization of systems\mycite{Hirokazu_PRL_2013, Aidelsburger_PRL_2013, Jotzu_Nature_2014}  with nontrivial topology. Together with the ability to engineer systems on a lattice scale\mycite{Bakr_Nature_2009,Wenz_Science_2013, Murmann_PRL_2015}, these developments usher in unprecedented possibilities. 

Quantum impurity problems  which provided many key concepts and ideas influencing all of physics\mycite{Wilson_RMP_1975, Hewson_Book_1997, HRK_PRB_1980, HRK_PRB_II_1980} has seen a recent resurgence e.g., from probing systems near a quantum critical point\mycite{Argha_PRB_2010, Dhochak_PRL_2010}, and even in numerical solution techniques\mycite{Bulla_RMP_2008, Georges_RMP_1996, Gull_RMP_2011}. This immutable importance of impurity problems has motivated several works\mycite{Bauer_PRL_2013, Nishida_PRL_2013, Falco_PRL_2004, Kuzmenko_PRB_2015} on using cold atom systems to study these. In this context, the new developments with synthetic gauge fields discussed above provide a new direction apropos physics of quantum impurities in these systems. 

A uniform non-Abelian ($SU(2)$) gauge field produces a generalized Rashba  spin-orbit coupling (RSOC) on the motion of spin-$1/2$ particles. There are several proposals (and lab realizations) \mycite{Goldman_RPP_2014}  for obtaining RSOC that produces spin-momentum-locking in one\mycite{Lin_PRL_2009, Cheuk_PRL_2012, Wang_PRL_2012}, two\mycite{Liu_PRL_2009,Lin_NatPhys_2011, Brandon_PRL_2013} and even three \mycite{Brandon_PRL_2012} spatial dimensions. The physics of quantum impurities in Fermi systems with RSOC has open questions: Are there new features when a quantum impurity hybridizes with a gas of RSOC fermions?  Is there a Kondo effect, and if so does it possess any unique features?   As will become evident in this paper, the answer to these questions is in the affirmative, and indeed there is new physics not yet uncovered in earlier work\mycite{Isaev_PRB_2012, Zarea_PRL_2012, Zitko_PRB_2011, Feng_JPCM_2011, Yanagisawa_JPSJ_2012, Malecki_JSP_2007, Chen_arXiv_2015}. 

Here we study a correlated quantum impurity (interaction scale $U$) which hybridizes (hybridization scale $V$) with a RSOC (strength $\lambda$) Fermi gas with an interparticle separation $k_F^{-1}$ (density $\sim k_F^3$, Fermi energy $E_F \sim k_F^2)$. When $\lambda/k_F \gtrsim 1$, we find that the impurity develops a {\em fractional local moment} (fraction is 2/3 for the 3D RSOC) for $U$ larger than a critical value. The moment couples  {\em antiferromagnetically} with the Fermi gas, and forms a Kondo like ground state. Quite remarkably, the resulting Kondo temperature $T_K$ is large -- a significant fraction of the Fermi energy -- and indeed can be increased with increasing RSOC ($T_K \sim \lambda^{4/3}$). We establish these results using a variety of methods from mean-field theory, variational ground state, and quantum Monte Carlo  numerics. Our analysis also demonstrates the physics behind the formation of the fractional local moment and provides a recipe to control its value. We discuss the experimental realization of these results in a cold atoms system, and also touch upon their relevance in strongly spin-orbit coupled condensed matter systems such as oxide interfaces.

\begin{figure}
  \includegraphics[width=\myfigwidth]{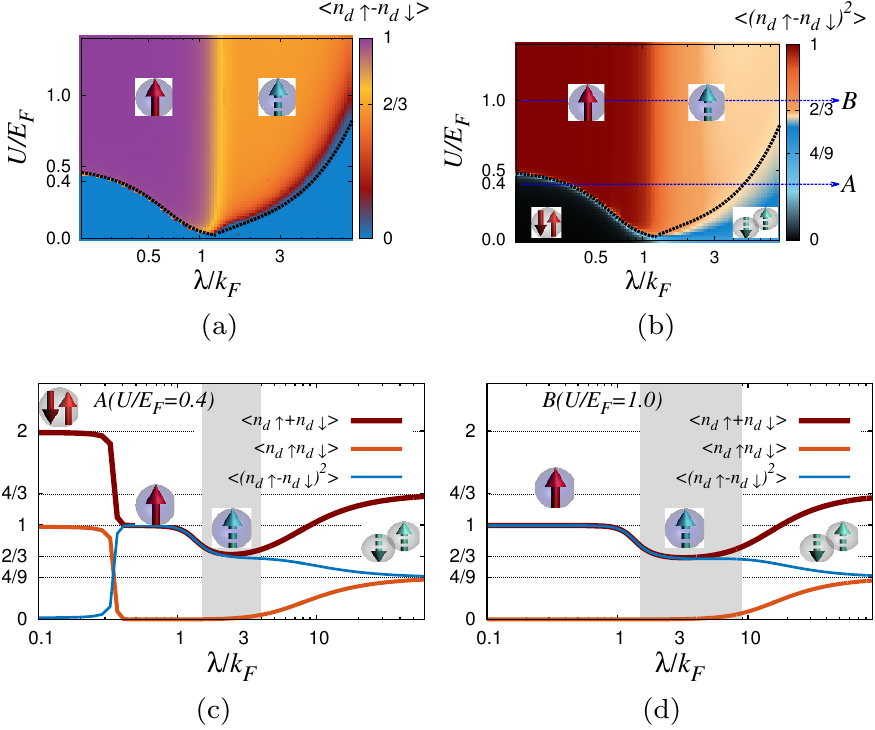}
  \caption{{\bf Ground state in $U$-$\lambda$ plane:} Results for $V/E_F^{1/4}=0.1$, $\varepsilon_d =\mu(\lambda)/2$. {\bf (a)} Impurity moment $M=\mean{n_{d\uparrow}-n_{d\downarrow}}$ in the Hartree-Fock (HF) ground state. {\bf (b)} Size of impurity moment $S^2_z=\mean{(n_{d\uparrow}-n_{d\downarrow})^2} $  in the variational ground state. {\bf (c)} and {\bf (d)} Results along slices $A (U/E_F=0.4)$ and $B(U/E_F=1.0)$ shown in {\bf (b)}.  Both HF and variational ground states show a {\em fractional} local moment (shown schematically by the broken vector) of $2/3$ for $\lambda/k_F \gtrsim 1$, and $U/E_F$ larger than a $\lambda$-dependent critical value shown by dashed line in {\bf (a)} and {\bf (b)}.}
\mylabel{fig:GS}
\end{figure}

\paratitle{Formulation} Consider a gas of two component (``spin-$\half$'') fermions in 3D with density $n_0  \equiv \frac{k_F^3}{3 \pi^2}$ with an associated energy scale $E_F = \frac{k^2_F}{2}$ {(Here and henceforth $\hbar$ and fermion mass are set to unity.)}. In the presence of RSOC induced by a non-Abelian gauge field, the spin of the fermions is locked to their momentum $\bk$ resulting in ``helicity'' $\alpha = \pm 1$ states. In terms of fermion operators $c^\dagger_{\bk \alpha}$, the Hamiltonian is $H_c = \sum_{\bk \alpha} (\varepsilon_\alpha(\bk) - \mu) c^\dagger_{\bk \alpha} c_{\bk \alpha}$, where $\varepsilon_\alpha(\bk) = \frac{k^2}{2} - \alpha |\bk_\lambda|$ is the ``kinetic energy'' {(for calculational convenience, energy is shifted by $\lambda^2/2$ (see Supplementary Information))}, $\bk_\lambda = \lambda_x k_x \be_x + \lambda_y k_y \be_y + \lambda_z k_z \be_z$. RSOC here, is described by $\blam = (\lambda_x,\lambda_y,\lambda_z) = \lambda \hat{\blam}$. The $\alpha=+1$ state has spin polarized along $\bk_\lambda$, while $\alpha=-1$ state has the spin opposite to $\bk_\lambda$, with $c^\dagger_{\bk \sigma} = \sum_\alpha f_\sigma^\alpha(\bk) c^\dagger_{\bk \alpha}$ with coefficients $f_\sigma^\alpha(\bk)$ determined by $\bk_\lambda$. Treating the Rashba-Fermi gas as a ``conduction bath'', we introduce an impurity state which we call the $d$-state following the usual terminology, which hybridizes with the gas. The impurity Hamiltonian is $H_d = \sum_{\sigma} (\tilde{\varepsilon}_d - \mu) n_{d \sigma} + U n_{d \uparrow} n_{d \downarrow}$ ($n_{d \sigma} = d^\dagger_\sigma d_\sigma$), where $\tilde{\varepsilon}_d$ is the ``bare'' impurity energy (see below), and $U$ is the crucial local repulsion between two fermions at the impurity site.  A second crucial aspect  is the local hybridization of the conduction fermions with the impurity state located at the origin of the 3D box of volume $\Omega$ given by $H_h =\frac{V}{\sqrt{\Omega}} \sum_{\bk \sigma} (c^\dagger_{\bk \sigma} d_\sigma + d^\dagger_\sigma c_{\bk \sigma})$. The Hamiltonian $H = H_c + H_d + H_h$ describes a cold atoms analog of an Anderson impurity problem\mycite{Anderson_PR_1961}. We focus on the case with 3D spin orbit coupling with $\lambda_x=\lambda_y=\lambda_z=\lambda$, the results of which are also applicable to other cases. Such an impurity system can be realized in an experiment by a combination of approaches described in refs.~\mycite{Brandon_PRL_2012} for the 3D RSOC, and \mycite{Bauer_PRL_2013} for the impurity.


The bath Fermi gas itself (without the impurity) undergoes changes due to the RSOC. For a given density $n_0$, increasing $\lambda$ causes a change in the topology of the Fermi surface\mycite{Jayantha_PRB2_2011}. Indeed for the 3D RSOC, this occurs at $\lambda_c = \frac{k_F}{\sqrt[3]{4}}$ and for $\lambda > \lambda_c$, the Fermi sea is a spherical annulus solely of $+$ helicity fermions. For $\lambda \ll \lambda_c$, the chemical potential varies as $\mu(\lambda)/E_F = 1 - \frac{1}{\sqrt[3]{2}} \left(\frac{\lambda}{\lambda_c} \right)^2$, and as $\frac{\mu(\lambda)}{E_F} = \frac{2^{8/3}}{9} \left(\frac{\lambda_c}{\lambda}\right)^{4}$ for $\lambda \gg \lambda_c$ . We next discuss the physics of a correlated impurity that hybridizes with this bath using various methods.

\paratitle{Ground State (Mean Field Theory)} The simplest approach that could reveal possible interesting physics arising from the impurity is the Hartree-Fock (HF) method\mycite{Anderson_PR_1961}. A broken (rotation) symmetry ground state is assumed, such that $M=\mean{n_{d \uparrow} - n_{d\downarrow}}$ is non-zero and self consistently determined. This calculation (and all others that we present below) requires an important technical input. Unlike the usual condensed matter problems where the bath has a well defined bandwidth, the 3D continuum fermions considered here do not. This leads to ultraviolet divergences (owing to the fermions at large momenta) requiring regularization. Our approach is to make the impurity energy $\tilde{\varepsilon}_d$ a bare parameter (see Supplementary Information for details), trading it for the physical value $\varepsilon_d$ via the relation $\varepsilon_d = \tilde{\varepsilon}_d - \frac{2V^2}{\Omega} \sum_{|\bk| \le \Lambda} \frac{1}{|\bk|^2}$, where $\Lambda$ is an ultraviolet cutoff. This procedure provides a route to make all interesting observables to be independent of cutoff $\Lambda$, not only for the HF approach but also for the others discussed below. \Fig{fig:GS}(a) shows ``magnetization'' $M$ of the impurity in the $U$-$\lambda$ plane, showing three distinct regimes. For any $\lambda$, $M$ vanishes when $U < U_c$ ($U_c(\lambda)$ is shown by the dashed line in \Fig{fig:GS}(a)). For $U > U_c$, $M \approx 1$ when $\lambda/k_F \lesssim 1$ consistent with known results\mycite{Anderson_PR_1961}. Most interestingly, for $\lambda/k_F  \gtrsim 1$ and $U > U_c$ we find that $M \approx 2/3$ motivating the more detailed investigations below.

\begin{figure} 
  \includegraphics[width=\myfigwidth]{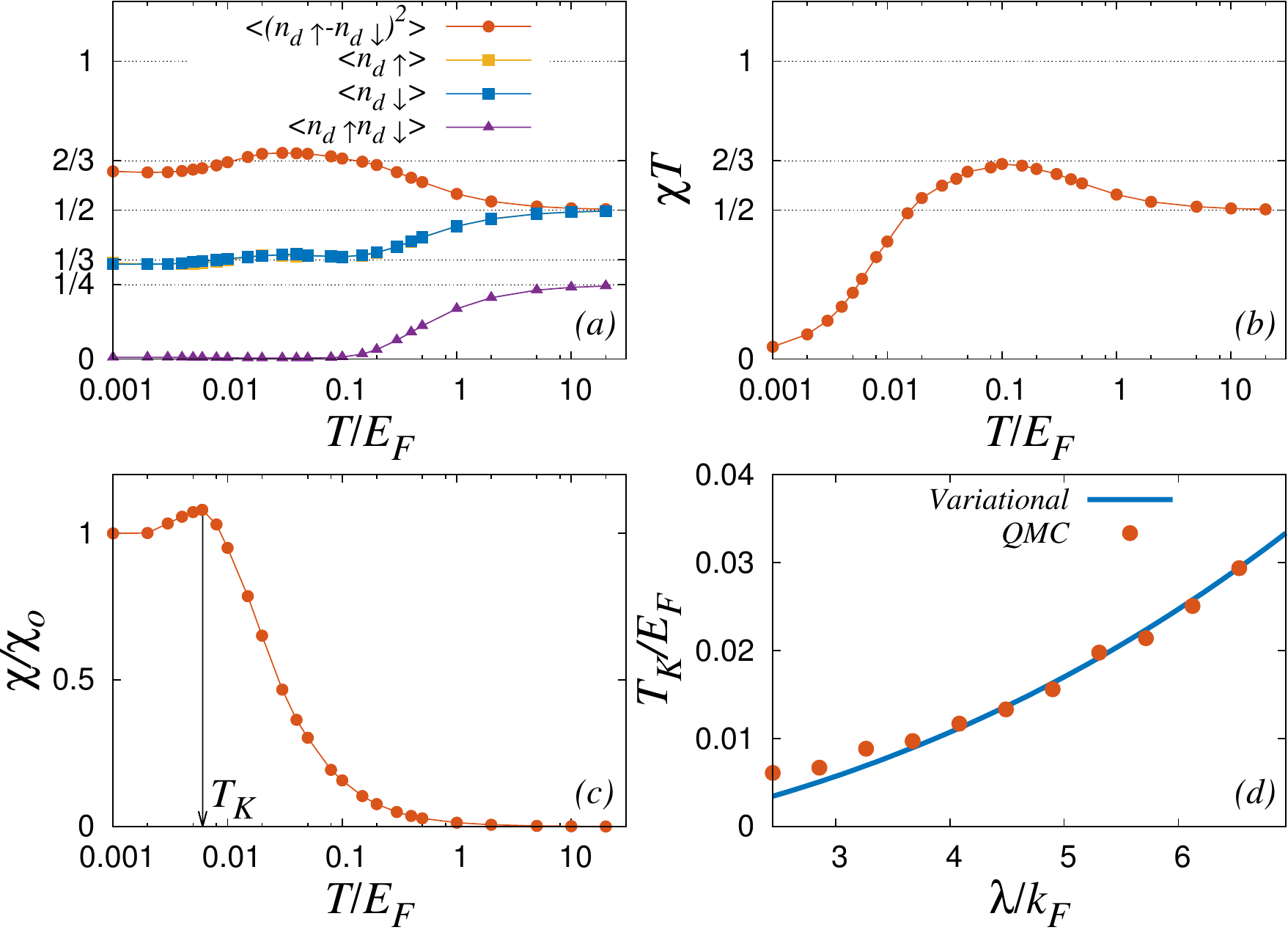}
  \caption{{\bf Finite $T$ physics:} QMC results for  $U/E_F=0.5, \lambda/k_F= \frac{5}{\sqrt{6}}$ and $V/E_F^{1/4}=0.1$ {\bf (a)} Impurity observables, {\bf (b)} and {\bf (c)} impurity magnetic susceptibility $\chi$ as a function of temperature $T$. $\chi_0$ in {\bf (c)} is the low temperature susceptibility. Kondo temperature $T_K$ is estimated from QMC results by the location of the peak in $\chi$ as shown in {\bf (c)}. {\bf (d)} Dependence of $T_K$ on $\lambda$($U/E_F=1.0$). Results {\bf (a) - (c)}  are obtained using $L=512$ imaginary time slices, while $L=128$ is used to obtain the $T_K$ for various values of $\lambda$ in {\bf (d)}. Sampling error bars are smaller than the symbol sizes.}
  \mylabel{fig:QMC}
\end{figure}

\paratitle{Ground State (Variational)} To obviate any artifacts due to the artificially broken symmetry of the HF calculation, we now construct a variational ground state (see e.~g., \mycite{Yosida_PR_1966}) with a ``rigid'' Fermi sea of bath fermions and two added particles whose spin-states are completely unbiased (see Supplementary Information for details). For the 3D RSOC, we find that the ground state for all $\lambda$ and $U$ is rotationally invariant with a zero total (spin+orbital) angular momentum  ($J=0$, singlet). The size of the impurity local moment, characterized by $S^2_z \equiv \mean{(n_{d \uparrow} - n_{d\downarrow})^2}$, depends on $\lambda$ and $U$ as seen from \Fig{fig:GS}(b), showing four
distinct ground states. (i) For $\lambda \lesssim k_F$ and $U < U_c$ ($U_c$ depends on $\lambda$, and is shown by a dashed line in \Fig{fig:GS}(b)), $S^2_z$ vanishes and the impurity is doubly occupied. (ii) For $\lambda \lesssim k_F$ and $U > U_c$, $S^2_z$ attains a value of unity corresponding to the Kondo ground state where the impurity has a well formed local moment that locks into a singlet with the bath fermions. Interestingly, in this regime of $\lambda$, $U_c$ falls with increasing $\lambda$, \ie small $\lambda$ aids the formation of the Kondo state(see also, \mycite{Zarea_PRL_2012}).
The other two states occur for $\lambda \gtrsim k_F$, where $U_c$ increases  with increasing $\lambda$. (iii) For $U > U_c$, we find a strongly correlated state (vanishing double occupancy) with a {\em fractional} local moment characterized by $ S_z^2 = 2/3$ ! (iv) For $U < U_c$ with $\lambda \gtrsim k_F$ , there is a intriguing new state with impurity occupancy of $4/3$, moment $4/9$, and a double occupancy $\mean{n_{d \uparrow} n_{d \downarrow}} = 4/9$. The crossovers between these states with increasing $\lambda$  are clearly demonstrated in \Fig{fig:GS}(c) which shows various quantities evolving with $\lambda$ for $U = 0.4 E_F$. Starting from a doubly occupied impurity, there is a crossover to the usual Kondo ground state with a singly occupied impurity with a unit local moment. There is a second crossover to the new kind of singlet state with a fractional local moment of $2/3$ (no double occupancy) in the regime $\lambda \sim k_F$. Finally, at a larger value of $\lambda$ there is a crossover to the other novel partially correlated singlet state of the type (iv) noted above. For  large $U$ ($>U_c(\lambda=0)$, see \Fig{fig:GS}(d)) the state starts off as a Kondo state at $\lambda=0$, crossing over to the two new states with a larger regime of a correlated fractional local moment state. Indeed, the HF results of the previous paragraph are consistent with those of the variational calculations(VC). The first excited state of the VC is a triplet state ($J=1$), the energy of this excited state compared with that of the singlet ground state gives an estimate of the Kondo scale $T_K$ which is discussed in greater detail below.

\begin{figure}
  \includegraphics[width=\myfigwidth]{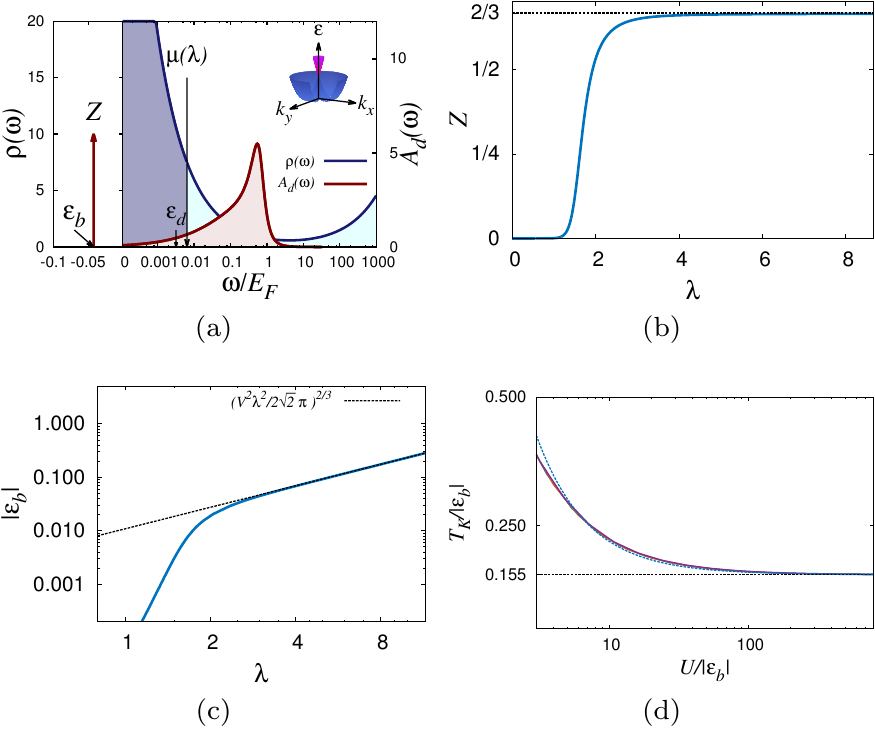}
  \caption{{\bf Fractional local moment and high $T_K$:} {\bf(a)} Density of states of the bath $\rho(\omega)$ with filled states up to $\mu(\lambda)$. The spectral function $A_d(\omega)$ of the $d$-state at $\varepsilon_d$ after hybridization with the bath is also shown. {\bf (b)} Weight($Z$) of the impurity $d$-state in the bound state, and {\bf (c)} Energy of the bound state $\varepsilon_b$, as function of $\lambda$. {\bf (d)} ``Universal'' Kondo $T_K$ scale as a function of $U$ estimated from the variational calculation where $\frac{T_K}{|\varepsilon_b|} \approx 0.155 + \frac{|\varepsilon_b|}{U}$  .
}
  \mylabel{fig:OneP}
\end{figure}

\paratitle{Finite Temperature (Quantum Monte Carlo)} Several natural questions arise including how the fractional local moment reveals itself at finite temperatures. We address this using the quantum Monte Carlo (QMC) method of Hirsch and Fye\mycite{Hirsch_PRL_1986} (see Supplementary Information for details). \Fig{fig:QMC}(a)-(c) shows the temperature dependent  results (including the impurity magnetic susceptibility $\chi$) obtained from QMC for a $\lambda$ and $U$ that possesses a fractional local moment in the ground state. Three temperature regimes are clearly seen. At high temperature $T \gg U$,  we have the ``free orbital regime''\mycite{HRK_PRB_1980, HRK_PRB_II_1980} where $T\chi(T) \approx \half$ (\Fig{fig:QMC}(b)), followed by a regime where $T \chi(T) \approx \frac{2}{3}$ at lower temperatures. At even lower temperatures (temperature scale $T_K$) there is a crossover to the Kondo state. The interesting aspects of these results is that the impurity local moment $S_z^2$ attains a value of $2/3$ in the same temperature regime where $T \chi(T) \sim \frac{2}{3}$ and remains so at low temperatures, even below the Kondo temperature $T_K$. This clearly indicates formation of a fractional local moment of $2/3$ at the impurity, and screening of the same by the bath fermions at lower temperatures. QMC also allows us to extract the Kondo temperature $T_K$ as shown in \Fig{fig:QMC}(c), and its dependence on $\lambda$ is shown in \Fig{fig:QMC}(d). The remarkable aspect is the large Kondo temperature scale that is a significant fraction of $E_F$, which interestingly increases with increasing $\lambda$ in the fractional local moment regime. Reassuringly, the energy scale obtained from the variational calculation also agrees with the QMC result (up to a factor of $\half$, $T_K^{QMC} \approx \frac{1}{2}T_K^{VC}$) as shown in \Fig{fig:QMC}(d).

\paratitle{Discussion} We now demonstrate that hybridization of the impurity with the Rashba-Fermi gas is behind the fractional local moment and the high $T_K$. In the absence of RSOC ($\lambda = 0$), the sole one-particle effect of hybridization on the impurity is to broaden its spectral function $A_d(\omega)$ from a Dirac delta at $\varepsilon_d$ to a Lorentzian of width $\Delta \sim V^2 \rho(\mu)$ where $\rho(\omega)$ ($\sim \sqrt{\omega}$ for $\lambda=0$) is the density of states of the bath. Matters take a different turn when $\lambda\ne0$ due to the infrared divergence of the density of states of the bath ($\rho(\omega) \sim \frac{\lambda^2}{\sqrt{\omega}}$ at near $\omega =0$, see \Fig{fig:OneP}(a)). A bound state appears for any $V$ for $\lambda \ne 0$, i.~e., the states $\{c^\dagger_{\bk \alpha}, d^\dagger_\sigma \}$ reorganize themselves into a set of scattering states created by $a^\dagger_{k m}$ and a {\em bound state} $b^\dagger_m$ ($k$ and $m$ are quantum numbers appropriate for the gauge field; for the 3D RSOC, $k=|\bk|$, $m$ is the $z$-projection of the total angular momentum $J=1/2$.). The weight $Z$ of the $d$-impurity state in the bound state $b$, i.e., $b^\dagger_{\half} = \sqrt{Z} d^\dagger_{\uparrow} + \sum_{\bk \alpha} B_{\bk \alpha} c^\dagger_{\bk \alpha}$, depends on $\lambda$ in a most interesting way. For a given $\varepsilon_d$ and $V$, $Z$ is vanishingly small for $\lambda$ smaller than a critical value (see \Fig{fig:OneP}(b)). For larger $\lambda$, $Z$ attains a constant value (of $\frac{2}{3}$ for the 3D RSOC) {\em independent} of $\lambda$. The energy of the bound state $\varepsilon_b$ also has interesting characteristics as shown in \Fig{fig:OneP}(c). For small $\lambda$, the binding energy is small and $\varepsilon_d$ dependent, while for large $\lambda$, $\varepsilon_b \approx - \left(\frac{V^2\lambda^2}{2 \sqrt{2} \pi} \right)^{2/3}$ and becomes independent of $\varepsilon_d$. 

\begin{figure} 
  \includegraphics[width=\myfigwidth]{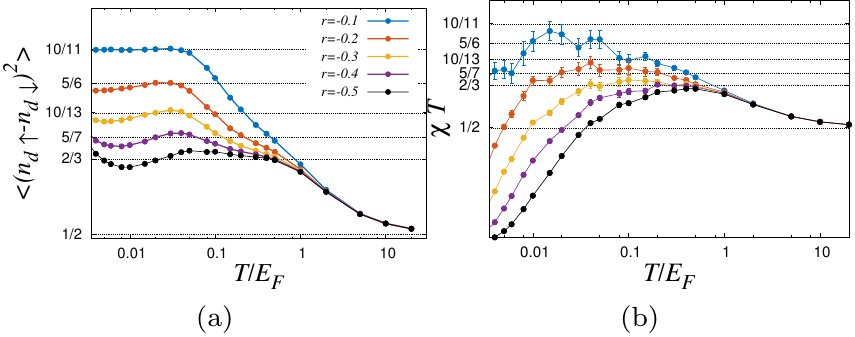}
  \caption{ {\bf Generic fractional local moments:} QMC results for an impurity hybridizing with a conduction band with $\rho(\omega)= \frac{1}{\pi^2}(\sqrt{2\omega}+\frac{\lambda}{\sqrt{2}}\frac{\omega^r}{\lambda^{2r}})$, $(V/E_F^{1/4}=0.1, \lambda/k_F = \frac{1}{\sqrt{2}}(\frac{10}{\sqrt{3}})^{\frac{2}{1-2r}}, U/E_F=2, \varepsilon_d = \mu_r(\lambda)/2, L=128)$. Impurity observables {\bf (a)} and susceptibility $\chi$ {\bf (b)}  as a function of temperature.
The values of $\varepsilon_b/E_F \approx -0.1$ for all cases.}
\mylabel{fig:BE}
\end{figure}

The one particle physics just discussed provides crucial clues to understanding of the physics as $U$ at the impurity site is increased. As is evident, the natural basis to understand the physics are the $b$-bound state and the $a$-scattering states. For small $\lambda$, the bound state has very little $d$-character and the physics is quite similar to the system without RSOC. The fall in $U_c$ seen in \Fig{fig:GS}(a,b) owes to the falling chemical potential of the gas for our choice of $\varepsilon_d = \mu(\lambda)/2$. At larger $\lambda$, the bound state $b$ is deep. Since the $b$ state has only a fraction $\sqrt{Z}$ of $d$ state, even a large $U$ on the $d$ state does {\em not} entirely forbid double occupancy of the $b$ state. Physically, the part of the $b$ state with $d$ character will ``feel'' a correlation energy $Z^2 U$, while the other part is uncorrelated. At large $U$, the ``$d$-part'' of $b$ will thus be singly occupied forming a fractional local moment. This argument provides an expression for the critical $U_c$ required to form a fractional local moment, as $U_c \sim \frac{1}{Z} |\varepsilon_b|$ and indeed matches (upto a multiplicative factor of $\approx2$) the result at large $\lambda$ shown in \Fig{fig:GS}(a,b). In fact, these observations also explain the regime of $U < U_c$ at large $\lambda$. Here the $b$ state is doubly occupied, and this corresponds to a $d$ occupancy of $2Z$, and $\mean{n_{d\uparrow}n_{d \downarrow}}= Z^2$ and $S_z^2 = 2Z(1-Z)$ all in agreement with results of \Fig{fig:GS}. Turning again to $U>U_c$, the origin of the high $T_K$ of the Kondo state formed by the fractional local moment can be understood from the variational calculation. As noted, the first excited state in VC is a triplet state made of a {\em singly occupied} $b$ state and a scattering state at the chemical potential, this is clearly a scale $\varepsilon_b$ above ground state with a fractional local moment and partial double occupancy of the $b$-state. Thus in the large $U$ limit we expect the Kondo scale $T_K$ to be proportional to $\varepsilon_b$ as indeed found by explicit calculation (see \Fig{fig:OneP}(d)). Indeed, this provides a route to obtain large Kondo temperatures as $T_K \sim \lambda^{4/3}$. Also note that the physics of the fractional local moment formation in this system is very different from that noted in ref.~\mycite{Vojta_EJP_2002} which occurs in a $sd$-system that has a {\em ferromagnetic} coupling to the bath.

A final puzzle: Why is $Z$ numerically equal to $\frac{2}{3}$? What controls this -- how can it be tuned? We show that $Z$ is {\em entirely determined by the exponent that characterizes the infrared divergence of the density of states.} Indeed, for a system with $\rho (\omega)=  \frac{1}{\pi^2}(\sqrt{2 \omega} + \frac{\lambda}{\sqrt{2}}\frac{\omega^r}{\lambda^{2r}})$ we show (see Supplementary Information) that $Z(r)=\frac{1}{1-r}$! We have performed QMC calculations with the impurity hybridizing to a bath with the given density of states, and indeed find the anticipated fractional local moments (see \Fig{fig:BE}(a)). We further see (\fig{fig:BE}(b)) that there are two distinct intermediate temperature regimes, $T_K \lesssim T \lesssim |\varepsilon_b|$ which is the  ``fractional local moment regime'' with $T \chi \approx Z(r)$, and the asymmetric local moment regime between $|\varepsilon_b| \lesssim T \lesssim U$ where $T \chi \approx \frac{2}{3}$. Interestingly, for the 3D RSOC the susceptibility alone cannot discern these two.

Experimental signatures of the fractional local moment formation in a cold atom experiment can be probed using radio-frequency(rf) spectroscopy\mycite{Ketterle_arXiv_2008}. The experiment will need a finite concentration of well separated quantum impurities, and the rf spectrum would show a well separated peak proportional to the concentration and the weight $Z$. In the condensed matter context, our results could also be useful in understanding experiments on low dimensional electron gases at oxide interfaces and surfaces \mycite{SantanderSyro_NatMat_2014, Gopinadhan_AEM_2015}.

\paratitle{Acknowledgements} The authors thank Diptiman Sen for discussions and suggestions. VBS thanks Michael Coey for a discussion on oxide interfaces.  AA acknowledges financial support from CSIR via SRF grant. VBS thanks DST and DAE for generous funding.


\bibliography{refAll} 

\clearpage



\appendix

\renewcommand{\appendixname}{}
\renewcommand{\thesection}{{S\arabic{section}}}
\renewcommand{\theequation}{\thesection.\arabic{equation}}
 
\setcounter{page}{1}
\setcounter{figure}{0}

\widetext

\centerline{\bf Supplementary Information}
\centerline{\bf for}
\centerline{\bf \titlename}
\centerline{by Adhip Agarwala and Vijay B.~Shenoy}
\author{Adhip Agarwala}
\email{adhip@physics.iisc.ernet.in}
\author{Vijay B. Shenoy}
\email{shenoy@physics.iisc.ernet.in}
\affiliation{Centre for Condensed Matter Theory, Department of Physics, Indian Institute of Science, Bangalore 560 012, India}

\section{Rashba-Anderson Hamiltonian}

The complete Hamiltonian $H=H_c+H_d+H_h$, as described in the main text is given by, 
\beq{\mylabel{eqn:Hamiltonian}}
\begin{split}
H = \sum_{\bk \alpha} \varepsilon_\alpha(\bk) c^\dagger_{\bk \alpha} c_{\bk \alpha} + \sum_{\sigma}{\tilde{\varepsilon}_{d}d_{\sigma}^{\dagger}d_{\sigma}} +  U n_{d \uparrow}n_{d \downarrow} + \\
\frac{V}{\sqrt{\Omega}}  \sum_{\bk, \sigma,\alpha}({f^{\alpha}_{\sigma}(\bk)}^*d_{\sigma}^{\dagger}c_{\bk \alpha} + {f^\alpha_{\sigma}(\bk)}c_{\bk \alpha}^{\dagger}d_{\sigma}).
\end{split}
\eeq
Here, ${f^{\alpha}_\sigma(\bk)}$ = $\langle \bk \alpha|\bk \sigma \rangle$ and $|\bk \alpha \rangle$ and $| \bk \sigma \rangle$ are the eigenkets of the Rashba spin-orbit coupled(RSOC) and the non-RSOC Fermi gas respectively. Explicitly $\langle\bk\sigma=\uparrow|\bk{\alpha=1}\rangle=\cos(\frac{\theta}{2})$,  $\langle\bk\sigma=\downarrow|\bk{\alpha=1}\rangle=\sin(\frac{\theta}{2})e^{i\phi}$, $\langle\bk\sigma=\uparrow|\bk{\alpha=-1}\rangle=-\sin(\frac{\theta}{2})$ and $\langle\bk\sigma=\downarrow|\bk{\alpha=-1}\rangle=\cos(\frac{\theta}{2})e^{i\phi}$, where $\theta$ and $\phi$ are the polar and azimuthal angles made by $\bk$ in spherical polar coordinates. In presence of RSOC (of strength $\lambda$) the two helicity bands($\alpha=\pm1)$ have the following dispersion (adding a constant energy shift of $\lambda^2/2$), 
\beq
\varepsilon_\alpha(\bk)= \varepsilon_\alpha(k)= (\frac{k}{\sqrt{2}} - \alpha \frac{\lambda}{\sqrt{2}})^2.
\eeq
$V$ is the strength of the hybridization of the impurity state $d$ with the conduction bath fermions $c_{\bk \alpha}$ and $\Omega$ is the volume. $\tilde{\varepsilon}_d$ is the impurity onsite energy and $U$ is the repulsive Hubbard interaction strength at the impurity site between two fermions.
The ``bath'' density of states, i.e. of the RSOC fermions is $\rho(\omega) = \frac{1}{\pi^2}(\frac{\lambda^2}{\sqrt{2\omega}} + \sqrt{2 \omega})$. Given a density of particles ($n_o=k_F^3/3\pi^2$), the chemical potential $\mu$ depends on $\lambda$ as \mycite{Jayantha_PRB2_2011}, 
\beq
\mylabel{eqn:chempot}
\sqrt{\frac{\mu}{E_F}}(\frac{3\lambda ^2}{k_F^2}+ \frac{\mu}{E_F}) = 1.
\eeq

\section{Ultraviolet Regularization and Impurity Spectral Function}
\mylabel{sec:imp}

The non-interacting impurity Green's function ($U=0$)  is given by, 
\beq
{\cal G}_{d \sigma}(\omega) = \frac{1}{(\omega- \tilde{\varepsilon}_{d}- \sum_{\bk,\alpha} \frac{V^2}{2\Omega} \frac{1}{(\omega- \varepsilon_{\alpha}(\bk))})}.
\eeq

The third term in the denominator of the above expression has an ultraviolet divergence. We describe the procedure of regularization mentioned in the main text. $\tilde{\varepsilon}_d$ is treated as a bare parameter and replaced by the corresponding physical parameter $\varepsilon_d$ using,
\beq
\varepsilon_d = \tilde{\varepsilon}_d - \frac{V^2}{\Omega} \sum_{|\bk| \le \Lambda} \frac{1}{|\bk^2/2|}=
\tilde{\varepsilon}_d - \frac{V^2 \Lambda}{\pi^2}.
\eeq
The regularized Green's function is, 
\beq
{\cal G}_{d \sigma}(\omega) = \frac{1}{(\omega- \varepsilon_{d}- (-\frac{V^2 \lambda^2}{2\sqrt{2}\pi\sqrt{-\omega}} +  \frac{V^2\sqrt{-\omega}}{\sqrt{2}\pi}))}.
\eeq

The corresponding impurity spectral function is,
\beq
A_d(\omega) = 2\pi Z \delta(\omega-\varepsilon_b) +   \frac{2 \left(\frac{\lambda ^2 V^2}{2 \sqrt{2} \pi  \sqrt{\omega }}+\frac{V^2 \sqrt{\omega }}{\pi  \sqrt{2}}\right)}{(\omega - \varepsilon_d)^2+\left(\frac{\lambda ^2 V^2}{2 \sqrt{2} \pi  \sqrt{\omega }}+\frac{V^2 \sqrt{\omega }}{\pi  \sqrt{2}}\right)^2}
\mylabel{eqn:Spec}
\eeq
where, $\frac{1}{2\pi}\int_{-\infty}^{\infty} A_d(\omega) d\omega= 1$. $\varepsilon_b$ is the pole of the Green's function and $Z$ is the weight of the $d$ state in the $b$ bound state. $Z$ is evaluated by the following procedure. For any impurity Green's function of the form ${\cal G}_{d\sigma}(\omega) = \frac{1}{f(\omega)}$, if $\varepsilon_b$ solves for the pole(\ie $f(\omega=\varepsilon_b)=0$), then $Z=\frac{1}{f'(\omega)}|_{\omega=\varepsilon_b}$.


\section{Hartree-Fock Method}

 Under the Hartree-Fock method(HF), the interaction term (see \eqn{eqn:Hamiltonian}) is treated as, 
\beq
U n_{d\uparrow} n_{d\downarrow} \rightarrow U (\langle n_{d \uparrow} \rangle n_{d \downarrow} + n_{d \uparrow} \langle n_{d \downarrow} \rangle - \langle n_{d \uparrow} \rangle \langle n_{d \downarrow} \rangle).
\eeq
The occupancy of $d$ state for both spin labels can now be self consistently found by solving,
\beq
\langle n_{d \sigma} \rangle = \int_{-\infty}^{\frac{\mu(\lambda)}{E_F}} \frac{-1}{\pi} \Im [ {\cal G}_{d\sigma}(\frac{\omega^+}{E_F},\frac{\varepsilon_d +  U \langle n_{d\bar{\sigma}} \rangle}{E_F}, \frac{V}{E_F^{1/4}},\frac{\lambda}{k_F})] d(\frac{\omega}{E_F}).
\eeq
This then allows us to find impurity moment $M=\langle n_{d\uparrow} - n_{d\downarrow}\rangle$ as a function of $U$ and $\lambda$, as is shown in the main text.

\section{Variational Calculation}

To build the variational calculation(VC), we first look at the resolution of identity in the non-RSOC basis,
\beq
1 = \frac{\Omega}{8\pi^3} ( \sum_{l,m,\sigma} \int_0^{\infty} dk k^2 |k,l,m,\sigma \rangle \langle k,l,m,\sigma| )
\eeq
where $k=|\bk|$, and $l,m$ and $\sigma=\pm1/2$ are the azimuthal, magnetic and the spin quantum numbers respectively. $|k,l,m,\sigma\rangle$ are therefore the free particle spherical wave states. Now $\lambda$ couples $l, m$ and $\sigma$ states to form $helicity$ states $(l, m, \sigma) \rightarrow (j, m_j, \alpha)$. Resolution of identity in this basis is,
\beq
1= \frac{\Omega}{8\pi^3} ( \sum_{j,m_j,\alpha} \int_0^{\infty} dk k^2 |k,j,m_j,\alpha \rangle \langle k,j,m_j,\alpha| )
\mylabel{ResolIden}
\eeq
where for any $k$,

\beq
|l=0,m=0,\uparrow\rangle= \frac{1}{\sqrt{2}}|j=\frac{1}{2},m_j=\frac{1}{2},\alpha=-1\rangle -  \frac{1}{\sqrt{2}}|j=\frac{1}{2},m_j=\frac{1}{2},\alpha=1\rangle 
\eeq
\beq
|l=0,m=0,\downarrow\rangle=\frac{1}{\sqrt{2}}|j=\frac{1}{2},m_j=-\frac{1}{2},\alpha=-1\rangle -  \frac{1}{\sqrt{2}}|j=\frac{1}{2},m_j=-\frac{1}{2},\alpha=1\rangle.
\eeq
The Hamiltonian $H$ can therefore be written as,
\beq{\mylabel{eqn:SphericalHamiltonian}}
\begin{split}
H  =\sum_{j,m_j,\alpha} \frac{\Omega}{8\pi^3} \int_{k}k^2dk \varepsilon_{\alpha}(k) |k,j,m_j,\alpha\rangle\langle k,j,m_j,\alpha| 
+ U |d,\sigma\rangle\langle d,\sigma|  |d, \bar{\sigma}\rangle\langle d,\bar{\sigma}|   
+ \sum_{\sigma} \tilde{\varepsilon}_d |d,\sigma\rangle\langle d,\sigma| \\ +  \sum_{\sigma,\alpha} V \frac{ \sqrt{\Omega} \sqrt{4\pi}}{8\pi^3}
( \times \int_{k}k^2 dk \frac{1}{\sqrt{2}}( \bar{\alpha}|k,j=1/2,m_j=\sigma,\alpha\rangle \langle d, \sigma| + h.c.) )
\end{split}
\eeq

Since $k \in (0,\infty)$, we transform  $k=\tan(\frac{\pi x}{2})$ such that $dk= jac(x) dx $ where, $jac(x)=\sec^2(\frac{\pi x}{2})\frac{\pi}{2}$. The $x$-interval $(0,1)$ is now further divided into discrete Gauss-Legendre points, $\int_0^{1} dx \rightarrow \sum_{i} wt(x_i)$, such that resolution of identity can be rewritten as,
\beq
1= \sum_{i,j,m_j,\alpha} g(x_i) |k(x_i),j,m_j,\alpha \rangle \langle k(x_i),j,m_j,\alpha|
\eeq
where, $g(x_i)=(\frac{\Omega}{8\pi^3}) wt(x_i) jac(x_i) k(x_i)^2 $. Defining $|\tilde{k}_i\rangle=|\tilde{k}(x_i)\rangle \equiv \sqrt{g(x_i)} |k(x_i)\rangle$ the complete discretized Hamiltonian is,

\beq
\begin{split}
H = \sum_{i,j,m_j,\alpha} \varepsilon_{\alpha}(k_i)| \tilde{k}_i,j, m_j, \alpha\rangle\langle \tilde{k}_i,j, m_j, \alpha| \\ + Un_{d\uparrow} n_{d\downarrow} + \sum_{\sigma} \tilde{\varepsilon}_d |d,{\sigma}\rangle \langle d, \sigma| \\+ V \sum_i \frac{1}{\sqrt{\Omega}} \sqrt{2\pi g(x_i)} (-|\tilde{k}_i , j=\frac{1}{2}, m_j=\frac{1}{2}, \alpha=1 \rangle \\ + |\tilde{k}_i, j=\frac{1}{2}, m_j=\frac{1}{2}, \alpha=-1 \rangle ) \langle d \uparrow|  \\ +  V\sum_i \frac{1}{\sqrt{\Omega}} \sqrt{2\pi g(x_i)} (-| \tilde{k}_i , j=\frac{1}{2}, m_j=-\frac{1}{2}, \alpha=1 \rangle \\+ | \tilde{k}_i , j=\frac{1}{2},m_j=-\frac{1}{2},  \alpha=-1 \rangle ) \langle d \downarrow|
\end{split}
\eeq

The system is numerically diagonalized in the non-interacting sector ($U=0$), where the regularization of $\tilde{\varepsilon}_d$ is included. A rigid Fermi sea is implemented by discarding states which have $\varepsilon_{\alpha}(\bk) < \mu(\lambda)$. The $U$ term of the Hamiltonian is further diagonalized in the two-particle sector using the product of one particle states. Various observables can then be calculated by taking expectation on the ground state wavefunction. Typically $\approx 10^4$ states in the two-particle sector may be necessary to find accurate solutions.

\section{Hirsch-Fye Quantum Monte Carlo}

Hirsch-Fye quantum Monte Carlo numerics are performed following \mycite{Hirsch_PRL_1986} where the susceptibility is obtained by,

\beq
\begin{split}
\chi = \int_0^{\beta} d\tau \langle [d^{\dagger}_{\uparrow}(\tau)d_{\downarrow}(\tau) + d^{\dagger}_{\downarrow}(\tau)d_{\uparrow}(\tau)] \times \\ [d^{\dagger}_{\uparrow}(0)d_{\downarrow}(0) + d^{\dagger}_{\downarrow}(0)d_{\uparrow}(0)] \rangle .
\end{split}
\eeq

The starting Green's function can be obtained from the non-interacting impurity spectral function (see \eqn{eqn:Spec}). Throughout the calculations, the chemical potential is kept fixed at its zero-temperature value ($\mu(T, \lambda)=\mu(\lambda))$. Our formulation can be readily used to obtain quantities of interest to experiments using realistic (temperature/system dependent) values of parameters.

\section{Infrared Divergence of Density of States determines {\it Z}} 

In order to understand the origin of $Z=2/3$, we construct conduction baths with infrared divergence in the density of states of the form, 
\beq
\rho (\omega)=  \frac{1}{\pi^2}(\sqrt{2 \omega} + \frac{\lambda}{\sqrt{2}}\frac{\omega^r}{\lambda^{2r}})
\eeq

The infrared divergence is characterized by the exponent $r$ $(-1<r<0)$. For a given density of particles $n_o$, one can obtain the dependence of $\mu$ on both $r$ and $\lambda$ (similar to \eqn{eqn:chempot}). The impurity Green's function and $Z$ is obtained as illustrated in \sect{sec:imp}. It is found that for large $\lambda$, $Z \rightarrow \frac{1}{1-r}$.  
This can be obtained analytically, by discarding the $\sim \sqrt{\omega}$ term in $\rho(\omega)$ and  considering $\rho(\omega)= \omega^r$ $(-1<r<0)$. The impurity Green's function in this case is given by, 
\beq
{\cal G}_d(\omega) = \frac{1}{\omega -\varepsilon_d - V^2 \pi (-\frac{1}{\omega})^{-r} \csc(\pi r)}
\eeq
with $Z=\frac{1}{1-r}$ for all values of $V(\ne 0)$ and $\varepsilon_d=0$.

\clearpage

\setcounter{page}{1}
\setcounter{figure}{0}

\end{document}